\documentclass[11pt]{article}
\usepackage[margin=1in]{geometry}

\usepackage{subcaption}
\usepackage{multirow}
\usepackage[round,authoryear]{natbib}

\usepackage[utf8]{inputenc} 
\usepackage[T1]{fontenc}    
\usepackage{url}            
\usepackage{booktabs}       
\usepackage{amsfonts}       
\usepackage{graphicx}
\usepackage{tikz}

\usepackage{csquotes}
\usepackage{times}
\usepackage{latexsym}
\usepackage{setspace}
\usepackage{enumitem}
\usepackage{makecell}
\usepackage{xcolor}
\usepackage{appendix}

\usepackage{hyperref}  
\usepackage{doi}

\usepackage{xr}
\newcommand\noii[1]{\vskip .02in\noindent {\em #1.}}
\newcommand\noib[1]{\vskip .02in\noindent {\bf #1}}

\newcommand\autocite{\citep}
\newcommand\textcite{\citet}
\renewcommand\cite{\citep}

\title{Causal Language in Observational Studies:\\ Sociocultural Backgrounds and Team Composition}

\author{
  Jun Wang\thanks{Independent Researcher, Syracuse, NY, USA} \\
  Bei Yu\thanks{School of Information Studies, Syracuse University, USA}
}

\date{}

\begin{document}
\maketitle


\abstract{
The use of causal language in observational studies has raised concerns about
overstatement in scientific communication. While some argue that such language
should be reserved for randomized controlled trials, others contend that
rigorous causal inference methods can justify causal claims in observational
research. Ideally, causal language should align with the strength of the underlying
evidence.
However, through the analysis of over 90,000 abstracts from observational studies
using computational linguistic and regression methods, we found that
causal language are more common in work by less
experienced authors, smaller research teams, male last authors, and researchers
from countries with higher uncertainty avoidance indices.
Our findings suggest that the use of causal language is not solely driven by the
strength of evidence, but also by the sociocultural backgrounds of authors and
their team composition. This work provides a new perspective for understanding
systematic variations in scientific communication and emphasizes the importance
of recognizing these human factors when evaluating scientific claims.
}

\section{Introduction}

Observational studies serve as crucial sources of evidence when randomized
controlled trials are infeasible or unethical.
However, due to the limitations of observational data,
whether to draw causal conclusions from these studies has been a subject of debate in the scientific community.
Drawing causal claims from observational evidence 
can raise concerns about overstated claims in the scientific literature, 
particularly when media coverage and press releases further amplify perceived certainty
\cite{sumner2014association,bleske2015causal}.
While some scholars argue that causal language should be restricted to evidence
from randomized controlled trials, as exemplified by medical journals like JAMA's editorial policies, 
others contend that rigorous causal inference methods can justify causal claims in observational research
\cite{pearl2018book,hernan2018c,dahabreh2024causal}.

To inform this ongoing debate, it is essential to understand what influences
researchers' choices to use causal language in their reporting. Ideally, such
language should reflect the strength of the underlying evidence. However,
sociology of science scholarship posits that knowledge production is shaped by
the context and identity of its producers \cite{haraway1988} and that scientific
writing is a social act \cite{Bazerman1983}. This framework views objectivity as
a collective property of scientific communities that emerges from shared
standards and openness to critique \cite{xia2024CitationSentiment}. Consequently, the
interpretation and communication of observational findings may reflect not only
the strength of the underlying evidence but also the sociocultural contexts in
which researchers operate. We therefore investigate whether sociocultural
factors, such as authors' cultural backgrounds, seniority and team composition,
influence how causal findings are framed in observational research.

Specifically, our goal was to examine whether characteristics of research teams and 
individual authors, including their country, gender, authorship position, team size, and
publication history, correlate with the use of causal language when presenting conclusions
in their structured abstracts.
To this end, we developed a
transformer-based computational linguistic algorithm that is capable of differentiating between causal and
correlational statements in biomedical domains. We applied it to over
91,933 observational study abstracts and analyzed the results using logistic
linear mixed-effects regression.
In this context, we identified observational studies using metadata assigned by
the NIH National Library of Medicine \cite{NIH_NLM_MeSH}, which defines an
observational study as ``a work that reports on the results of a clinical study
in which participants may receive diagnostic, therapeutic, or other types of interventions,
but the investigator does not assign participants to specific interventions (as in an interventional study).''

Our analysis controlled for potential confounders
including journal impact factor, study design, and 
a comprehensive set of over 1,400 Medical Subject Heading (MeSH) terms that
together capture a wide range of research topics, methodologies,
and key scientific concepts.
Additionally, to account for journal-specific effects related to
editorial policies and temporal trends, we modeled journal ISSN and publication
year as random effects. After adjusting for these factors, we found that causal
claims were more frequently used by less experienced authors, smaller teams,
male last authors, and authors from countries with higher uncertainty avoidance
indices \cite{hofstede2010}.
\section{Results}

\begin{figure*}[!t]
    \centering
    \begin{minipage}{0.28\textwidth}
        \centering
        \includegraphics[width=\textwidth, trim=0 0.25in 0 0, clip]{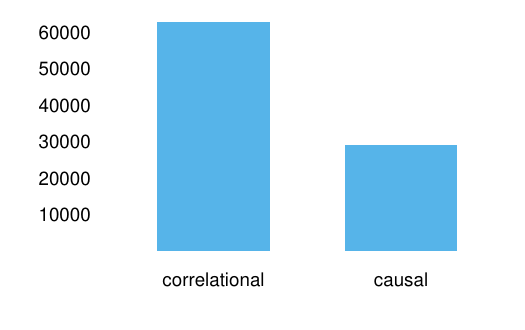} \\ (a) \\
        \includegraphics[width=\textwidth, trim=0 0.20in 0 0.05in, clip]{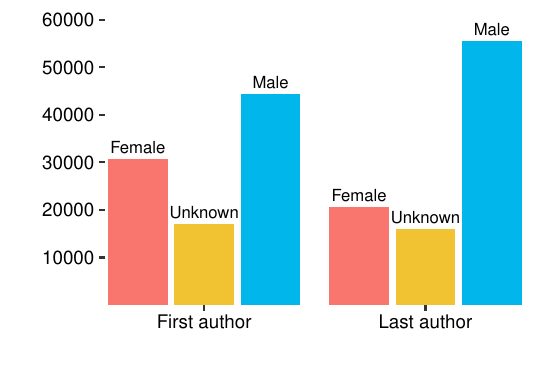} \\ (b) 
    \end{minipage}
      \hfill
    \begin{minipage}{0.71\textwidth}
        \centering
        \includegraphics[width=\textwidth, trim=0 0.38in 0 0, clip]{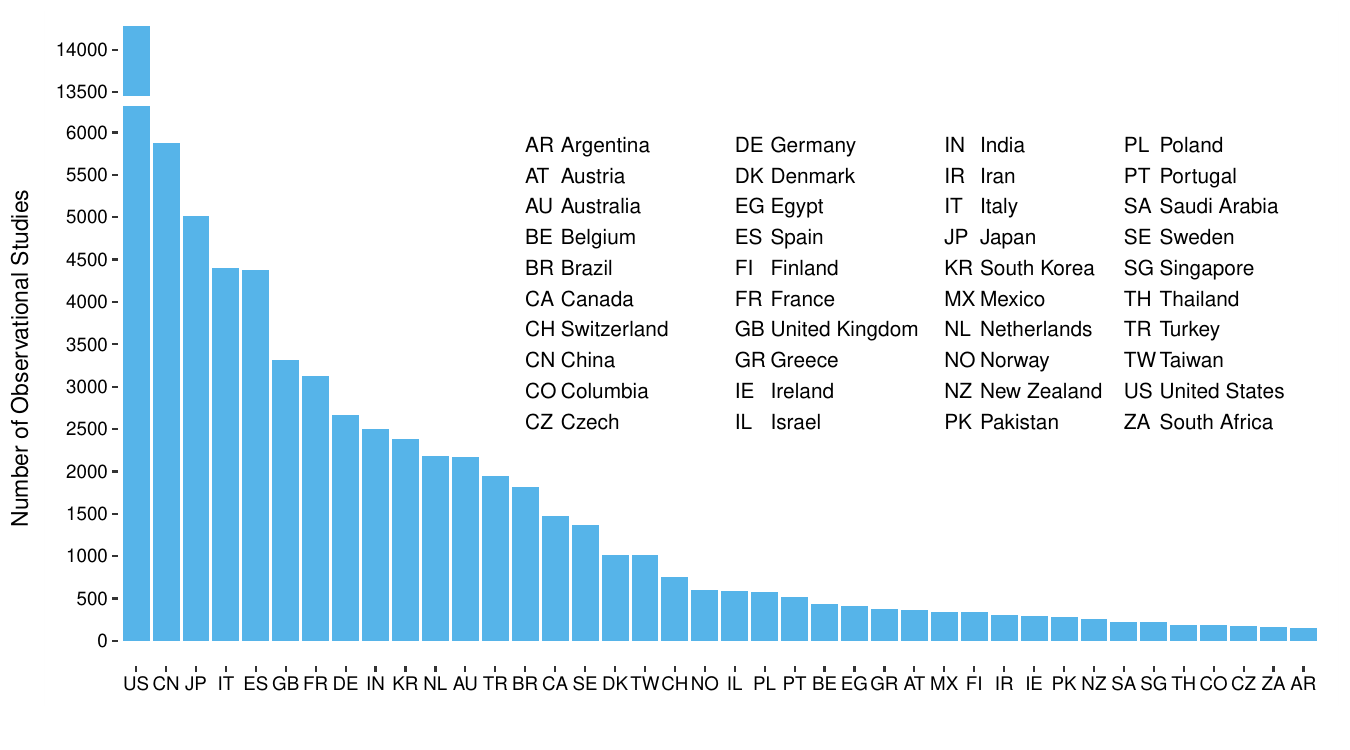} \\ (c)
    \end{minipage}

    \begin{minipage}{\textwidth}
        \centering
        \includegraphics[width=.48\textwidth]{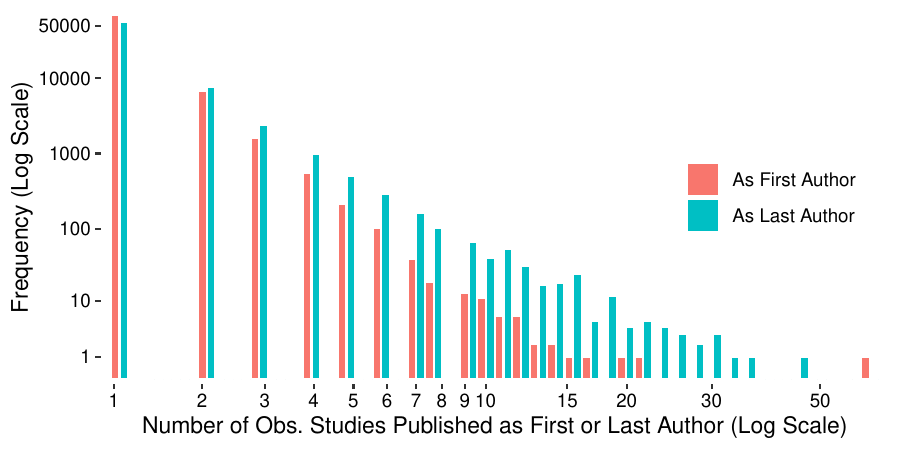}
        \;\;
        \includegraphics[width=.48\textwidth]{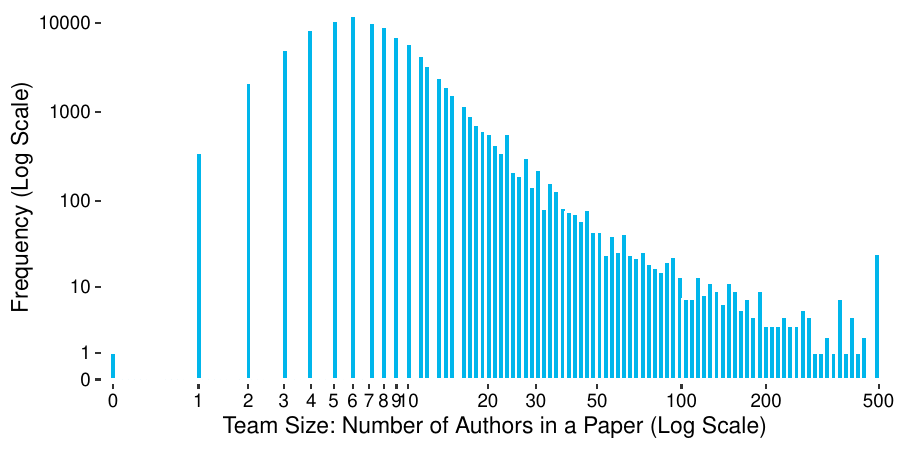} \\ 
    \end{minipage}

    \begin{minipage}{\textwidth}
      \begin{minipage}{0.48\textwidth}
          \centering
          (d)
      \end{minipage}
      \hfill
      \begin{minipage}{0.48\textwidth}
          \centering
          (e)
      \end{minipage}
    \end{minipage}
    \caption{Distribution of the dependent variable (causal vs correlational) along with the key variables of interest}
    \label{fig_data_summary}
\end{figure*}

%
\subsection*{Data Summary}
Fig.~\ref{fig_data_summary} 
presents the distribution of our dependent variable along with the key variables of interest.

\noii{(a) Dependent variable (causal or correlational)}
In our study, the dependent variable is defined as whether there is a causal statement in
the conclusion of an observational study abstract. We developed
a transformer-based algorithm to identify causal and
correlational statements in biomedical domains \cite{yu2019EMNLPCausalLanguage}.
Of the 91,933 observational study abstracts that contain causal or correlational statements, 
we found that 31.7\% had causal claims (Fig.~\ref{fig_data_summary}-a).
For comparison, a similar proportion (31\%) was reported by \cite{cofield2010use},
who manually analyzed 525 peer-reviewed papers in the fields of obesity and nutrition.

\noii{(b) Gender}
Fig.~\ref{fig_data_summary}-b illustrates the gender distribution by author position.
Following previous research on authorship in biomedical fields \cite{lerchenmueller2019gender},
we only consider first and last author,
as these positions in biomedical research typically represent the primary
contributors: first authors are usually those who contributed most substantially
to the writing, while last authors are the senior or principal investigators
overseeing the project. 

Author gender was inferred from given names, categorized as male, female, or
unknown.  The unknown category included cases where a given name was
missing (e.g., only initials provided) 
or where the gender prediction algorithm returned a low-confidence result. About
60\% of the unknown cases were due to the latter, a challenge particularly common
for names without clear gender markers, such as many Chinese and Korean names.

As shown in Fig.~\ref{fig_data_summary}-b, the male-to-female ratio for first
authors is 1.4:1, while for last authors, it increases to 2.7:1---nearly doubling
the ratio for first authors. This pattern, where male authors outnumber female
authors in both roles, aligns with previous findings on gender imbalances
in biomedical research publications \citep{ioannidis2023gender}, 
with a more noticeable difference observed among last authors.

\noii{(c) Country}
In this study, we consider only countries with at least 150 observational
studies in our dataset, totaling 40 countries (Fig.~\ref{fig_data_summary}-c).
These range from the United
States (14,278 papers) and China (5,871) to South Africa (169) and Argentina (158).
Studies from countries outside these 40, those 
with authors from multiple countries, or with missing or uncertain country
affiliations (see SI B.2), are grouped together and used as the reference
category in the regression analysis.

\noii{(d) Author publication experience}
Fig.~\ref{fig_data_summary}-d
displays the distribution of the number of observational studies
published by authors in either the first or last author position, with both axes
on a log scale. The x-axis represents the number of papers an author has
published in either role, while the y-axis indicates the frequency of authors
with that publication count. The distribution follows a heavy-tailed pattern,
where most authors have published only a few papers,
while a relatively small group of authors has a significantly higher publication count.

\noii{(e) Team size}
Fig.~\ref{fig_data_summary}-e
presents the distribution of team sizes,
measured by the number of authors per paper on a log scale for both the
x-axis and y-axis. The x-axis represents the number of authors per paper,
ranging from single-author papers
 to large collaborative teams exceeding 500
authors (as we can only retrieve up to 500 authors by combining PubMed and the Semantic Scholar data).
The y-axis denotes the frequency of papers with a given team size, also on a log scale.
The distribution exhibits a right-skewed pattern, indicating that most papers
have relatively small author teams, with the highest frequencies occurring for
papers with around 6 authors.

\begin{table}[!ht]
\centering

\begin{tabular}{@{}r@{ }l@{}}
\toprule

\begin{tabular}[t]{@{}l@{ }r@{ }l@{}l@{}}
\multicolumn{4}{@{}l}{\textbf{Author country}} \\
 New Zealand & -0.354 & (0.151) & $^{*}$ \\ [-.022in]
 United States & -0.308 & (0.027) & $^{***}$ \\ [-.022in]
 Norway & -0.298 & (0.101) & $^{**}$ \\ [-.022in]
 Denmark & -0.274 & (0.080) & $^{***}$ \\ [-.022in]
 Canada & -0.270 & (0.068) & $^{***}$ \\ [-.022in]
 South Africa & -0.186 & (0.180) & $^{}$ \\ [-.022in]
 Sweden & -0.162 & (0.068) & $^{*}$ \\ [-.022in]
 Israel & -0.137 & (0.099) & $^{}$ \\ [-.022in]
 Australia & -0.125 & (0.054) & $^{*}$ \\ [-.022in]
 United Kingdom & -0.119 & (0.045) & $^{**}$ \\ [-.022in]
 Finland & -0.088 & (0.130) & $^{}$ \\ [-.022in]
 Argentina & -0.086 & (0.184) & $^{}$ \\ [-.022in]
 Iran & -0.080 & (0.134) & $^{}$ \\ [-.022in]
 Japan & -0.068 & (0.040) & $^{}$ \\ [-.022in]
 Mexico & -0.042 & (0.127) & $^{}$ \\ [-.022in]
 Singapore & -0.038 & (0.158) & $^{}$ \\ [-.022in]
 South Korea & -0.023 & (0.054) & $^{}$ \\ [-.022in]
 Taiwan & -0.022 & (0.078) & $^{}$ \\ [-.022in]
 Saudi Arabia & -0.010 & (0.152) & $^{}$ \\ [-.022in]
 Turkey & 0.032 & (0.057) & $^{}$ \\ [-.022in]
 Egypt & 0.047 & (0.114) & $^{}$ \\ [-.022in]
 China & 0.050 & (0.037) & $^{}$ \\ [-.022in]
 France & 0.055 & (0.044) & $^{}$ \\ [-.022in]
 Ireland & 0.062 & (0.132) & $^{}$ \\ [-.022in]
 Brazil & 0.072 & (0.059) & $^{}$ \\ [-.022in]
 Portugal & 0.117 & (0.101) & $^{}$ \\ [-.022in]
 Netherlands & 0.122 & (0.051) & $^{*}$ \\ [-.022in]
 Spain & 0.141 & (0.039) & $^{***}$ \\ [-.022in]
 Columbia & 0.174 & (0.161) & $^{}$ \\ [-.022in]
 Czech & 0.179 & (0.165) & $^{}$ \\ [-.022in]
 Switzerland & 0.184 & (0.083) & $^{*}$ \\ [-.022in]
 Thailand & 0.198 & (0.161) & $^{}$ \\ [-.022in]
 India & 0.202 & (0.051) & $^{***}$ \\ [-.022in]
 Belgium & 0.215 & (0.108) & $^{*}$ \\ [-.022in]
 Germany & 0.225 & (0.046) & $^{***}$ \\ [-.022in]
 Greece & 0.250 & (0.117) & $^{*}$ \\ [-.022in]
 Poland & 0.300 & (0.096) & $^{**}$ \\ [-.022in]
 Italy & 0.329 & (0.038) & $^{***}$ \\ [-.022in]
 Austria & 0.346 & (0.123) & $^{**}$ \\ [-.022in]
 Pakistan & 0.350 & (0.142) & $^{*}$ \\ [-.022in] 

\end{tabular} 
& 

\begin{tabular}[t]{@{}l@{ }r@{ }l@{}l@{}}

\multicolumn{4}{@{}l}{\textbf{First author gender}} \\
  Men & 0.010 & (0.018) & $^{}$ \\
  Unknown &  0.033 & (0.025) & $^{}$ \\[.05in]

\multicolumn{4}{@{}l}{\textbf{Last author gender}} \\
  Men & 0.057 & (0.020) & $^{**}$ \\
  Unknown &  0.032 & (0.028) & $^{}$ \\[.05in]

\multicolumn{4}{@{}l}{\textbf{Obs. studies published} (log-scaled)} \\
 as first author & -0.119 & (0.028) & $^{***}$ \\
 as last author &   -0.167 & (0.018) & $^{***}$ \\[.03in]

\multicolumn{4}{@{}l}{\textbf{Team size} (log-scaled)} \\
 Num co-authors & -0.089 & (0.017) & $^{***}$ \\ [.03in]

\midrule 
\multicolumn{4}{@{}l}{\textbf{Control Variables}} \\ [.05in]

\multicolumn{4}{@{}l}{\textbf{Journal} (log-scaled)} \\
 SciMago journal rank & -0.091 & (0.032) & $^{**}$\\ 
 Obs. studies pub. & -0.042 & (0.009) & $^{***}$ \\ [.05in]

\multicolumn{4}{@{}l}{\textbf{Study design}} \\
  Cross-Sectional & -0.339 & (0.028) & $^{***}$ \\
  Case-Control & -0.278 & (0.043) & $^{***}$ \\
  Longitudinal & -0.161 & (0.040) & $^{***}$ \\
  Cohort & -0.161 & (0.028) & $^{***}$ \\
  Retrospective & -0.022 & (0.020) & $^{}$ \\
  Prospective & 0.026 & (0.019) & $^{}$ \\
  Follow-Up & 0.063 & (0.030) & $^{*}$ \\ 
[.05in]

\multicolumn{4}{@{}l}{\textbf{Other terms beyond study design}} \\
\multicolumn{4}{l}{ 1400+ MeSH terms (see SI D), each} \\ 
\multicolumn{4}{l}{ associated with 100 or more papers} \\ 
[.05in]

\midrule 
\multicolumn{4}{@{}l}{\textbf{Random Effects}} \\ [.05in]
 \multicolumn{4}{l}{Journal ISSN} \\
 \multicolumn{4}{l}{Publication year} \\
 \multicolumn{4}{l}{Conclusion length (num of sentences)} \\ [.03in]

\end{tabular} \\

\bottomrule
\end{tabular}

\caption{
Estimated effects of author country, gender, authorship position, team size, and writing experience on the likelihood of using causal language in the conclusion section of observational study abstracts.
 For author gender, the reference category is Female;
for author country, it is Others; 
and for study design and other MeSH terms, the reference category is the absence of the term.
\footnotesize{$^{***}p<0.001$; $^{**}p<0.01$; $^{*}p<0.05$}.
 }
\label{table__overall_effects}
\end{table}

\begin{figure}[!ht]
\centering
\includegraphics[width=.7\textwidth]{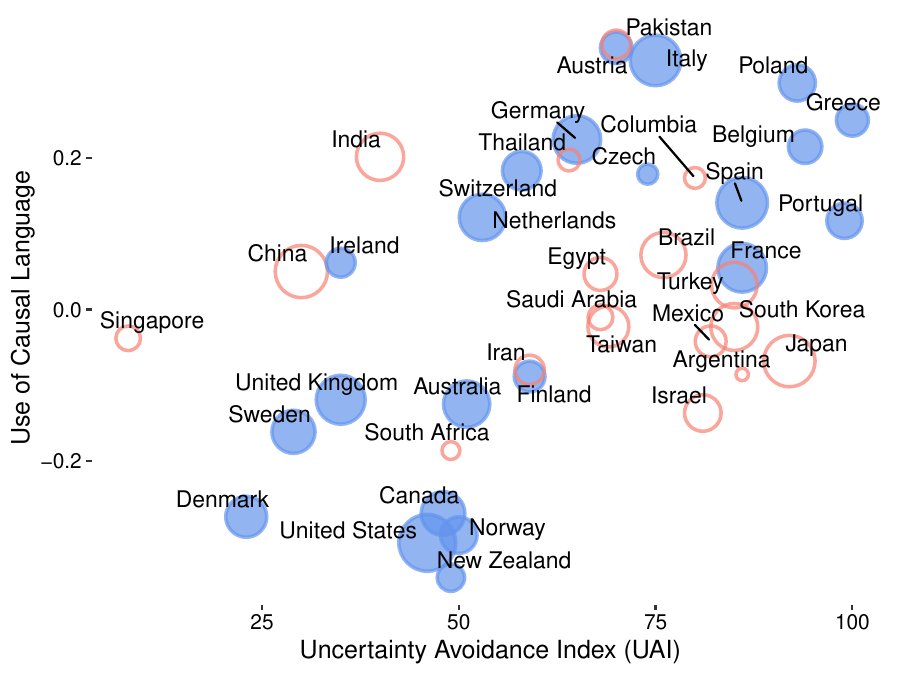}
\caption{
   Authors from countries with higher uncertainty avoidance index scores---reflecting a
    greater cultural preference for certainty---tend to use causal language more frequently.
    The correlation is moderate across all 40 countries (Pearson's $r=0.45, p < 0.01$),
    and strengthens significantly among 22 Western cultural countries ($r = 0.70, p < 0.001$).
    The size of the circles corresponds to the log scale of the number of observational studies published under the country’s name.
    }
\label{fig__uncertainty_vs_causal_all}
\end{figure}

\subsection*{Country and uncertainty avoidance culture}

Examining the 40 countries with at least 150 observational studies in our dataset,
 we find that geographic location appears to
correlate with authors’ tendency to use causal language
(Table~\ref{table__overall_effects}). Authors from North America (USA and
Canada) and Scandinavian countries (Denmark, Norway, and Sweden) were among the
least likely to use causal language—note that although Finland is a Nordic
country like its Scandinavian neighbors, its authors’ use of causal language
showed a much different pattern. In contrast, authors from Central and Southern
European countries—including Austria, Germany, Italy, Greece,
and Poland—demonstrated a significantly higher likelihood of
doing so.

This geographical variation raises the question of whether deeper cultural
factors might underlie cross-country differences in causal language use. To
explore this possibility, we focused on national-level cultural metrics,
specifically the uncertainty avoidance index (UAI)---a component of Hofstede’s
cultural dimensions that reflects a culture’s tolerance for ambiguity and
uncertainty \cite{hofstede2010}---and examined its relationship with countries’
use of causal language.

We visualized this relationship in Figure~\ref{fig__uncertainty_vs_causal_all},
which plots countries’ average use of causal language against their UAI scores.
Each country is represented by a circle log-scaled to
the number of observational studies in our dataset, with filled markers
indicating Western cultural countries. A clear positive trend emerges: countries
with higher UAI scores---such as Germany, Italy, Poland, and Greece---tend to use
causal language more frequently, while countries with lower UAI scores—such as
the United States, Canada, and the Scandinavian nations—use it less often. This
pattern is particularly strong among Western cultural countries, for which the
correlation is substantial (Pearson’s $r$ = 0.70), compared to a
moderate overall correlation across all 40 countries ($r$ = 0.45).
These findings suggest that cultural attitudes toward uncertainty may
systematically influence how researchers frame their conclusions in
observational work.

\subsection*{Other Key Variables of Interest}

\noib{Author Gender and Authorship Position.}
Our analysis reveals gender-related patterns in the use of causal language,
particularly among last authors. While first authors—often junior
researchers—showed no significant gender differences, male last authors used
causal language significantly more frequently than their female counterparts 
($\beta=0.057, p<0.01$).
Given that last authorship typically reflects seniority and leadership in
biomedical research, these findings suggest that gender differences in causal
framing may emerge more prominently in positions with greater influence over
study interpretation and presentation.

\noib{Author Publication Experience.}
Greater publication experience was significantly associated with reduced use of
causal language. Each log-unit increase in publication count corresponded to a
decrease of $\beta = -0.119$ ($p < 0.001$) for first authors and $\beta =
-0.167$ ($p < 0.001$) for last authors. This translates to approximately 8\%
and 11\% lower odds of using causal language, respectively, for each doubling
of publication count. The stronger effect among last authors suggests 
senior researchers may play a key role in framing causal claims,
consistent with the gender differences identified in our analysis.

\noib{Team Size.}
Our analysis also demonstrates that studies conducted by larger teams were less inclined to use causal language,
with a coefficient of $\beta=-0.089$ ($p < 0.001$) per log-unit increase in the number of co-authors.
The presence of diverse perspectives in larger collaborative groups
may encourage more nuanced discussions and negotiations, which in turn may contribute to a
more restrained interpretation of observational study results.
This finding suggests that collective authorship potentially serves as a
moderating influence on causal inference in observational research.

%
\subsection*{Control Variables} 
Our analysis of control variables shows patterns that align with conventional
understandings in scientific communication:

\noib {\em Journal Rank and Publication Volume.}
Higher-ranked journals, as measured by the SciMago Journal Rank, and journals that
publish a larger volume of observational studies were significantly less likely
to feature causal language. This pattern is consistent with editorial norms that
emphasize cautious interpretation of non-experimental findings. One possible
explanation is a filtering effect: top-tier journals, such as JAMA, may enforce
stricter language guidelines during peer review and editorial screening, leading
to more conservative reporting of observational results. This suggests that
journal-level policies and review standards may play a critical role in shaping
how authors frame causal claims.

\noib {\em Study Design.}
The likelihood of using causal language varied systematically by study design,
increasing in alignment with the established hierarchy of evidence in the
Evidence-Based Medicine Pyramid \cite{murad2016new}. On average, cross-sectional
studies were least likely to use causal terms, followed by case-control,
longitudinal, cohort, retrospective, prospective, and follow-up studies. This
gradient suggests that authors may be adjusting their language to match
perceived study rigor---using more cautious language in weaker designs and
expressing stronger causal interpretations as the design affords more robust
inference.

\noib {\em Over 1400 MeSH Terms.}
To account for topic-specific variation, our regression model also included
1,400+ Medical Subject Headings (MeSH) that appeared in at least 100 studies each. 
These terms cover a wide array of topics, methods, concepts, and clinical areas relevant to observational research.
The detailed effects of individual MeSH terms
on the use of causal language are given in SI D, offering additional insight into how the content and
focus of a study may influence authors’ linguistic choices---beyond the above study design.

\section{Discussion}

This study provides a new perspective on how human factors, such as authors'
sociocultural backgrounds and team composition, might influence the use of causal
language in observational studies. One key finding is the association
between a country's level of uncertainty avoidance \cite{hofstede2010} and the
tendency to use causal language; authors from countries with higher
uncertainty avoidance scores are more inclined to present their findings
as causal. This observation is generally in line with existing literature on
cultural influences in communication and decision-making, suggesting that
individuals in high uncertainty avoidance cultures may lean toward more
definitive conclusions and causal explanations as a way to reduce ambiguity and
convey greater certainty.

This cultural tendency toward certainty in causal framing may also manifest more
subtly in linguistic choices beyond causal claims themselves. Related patterns
are observed in the use of epistemic modal markers, which reflect how certainty
is linguistically expressed: for example, French researchers, representing a high
uncertainty avoidance context, use fewer expressions of uncertainty compared to
English or Norwegian researchers \cite{vold2006}, and German texts favor
stronger modal expressions over possibility markers common in English
\cite{kranich2011}. While not the primary focus here, these findings support the
broader link between cultural attitudes toward uncertainty and how scientific
claims are framed.

Another key finding in our study relates to gender and authorship roles. While
no significant gender differences were observed among first authors, our results
suggest that male last authors, typically senior researchers, tend to use causal
language more frequently than their female counterparts. This pattern may
reflect the combined influence of academic hierarchies and linguistic
preferences. As senior authors often shape the overall framing of a study, their
linguistic choices could influence how conclusions are presented.

In particular, male authors, who are more likely to use linguistic boosters
\cite{tannen1995,nasri2018projecting}, may frame study conclusions more
assertively and favor causal language. In contrast, female authors, who tend to
use more hedging language, may adopt a more cautious tone. A comprehensive
meta-analysis by Leaper and Robnett \citeyearpar{leaper2011} supports this distinction, showing
that women are somewhat more likely than men to use tentative language,
including hedges. These results suggest that gender
differences in expressing certainty may influence not only the use of hedges or
boosters, but also the broader tendency to frame research conclusions as causal
or correlational.

Our analysis also suggests that larger author teams may adopt a more cautious
approach when drawing causal conclusions from observational studies. This
tendency could reflect the benefits of collective deliberation: with more
contributors (often bringing diverse backgrounds and expertise), there are more
opportunities to scrutinize claims and refine the language. Prior work in
academic discourse supports this idea. Hyland argues that scientific writing is
not merely about reporting findings, but about negotiating meaning with an
audience through shared rhetorical norms \cite{hyland2004disciplinary}. In
collaborative writing, this negotiation occurs not only between writers and
readers but also among co-authors themselves. As a result, larger teams may
produce more rhetorically balanced manuscripts, as internal dialogue fosters
careful framing and reduces the likelihood of overstated claims.

Additionally, our findings also suggest that experienced authors, particularly those in senior
roles, tend to present their conclusions more cautiously. This may reflect a
deeper awareness of research limitations as well as a stronger grasp of
disciplinary writing norms. As Hyland notes, effective scientific communication
involves aligning language with the expectations of the research community
\cite{hyland2009writing}. Less experienced researchers, by contrast, may not yet
be fully attuned to these conventions, making them more likely to use causal
language in ways that diverge from field standards. At the same time,
alternative explanations warrant consideration. Junior authors may exhibit greater familiarity with 
contemporary statistical techniques and therefore feel more justified
in making causal claims, or they may be less constrained by traditional
rhetorical norms. Nonetheless, the consistent association between publication
experience and more cautious language use points to author seniority as an
important moderating factor in how research findings are framed.

One limitation of our study is that we did not assess the intrinsic causal
strength of the evidence presented in the observational studies we analyzed.
While our focus was on the linguistic framing of causal language, a more
comprehensive approach would involve quantifying causal strength alongside
linguistic analysis. Developing methods to estimate and incorporate causal
strength could provide a richer understanding of how the quality and robustness
of evidence influence the framing of causal claims in scientific writing.
\section{Materials and Methods}

\noib{Data.}
We queried the PubMed database using the search term ``Observational
Study[Publication Type]'' and retrieved over 180,000 papers, most of which were
published in 2014 or later following the introduction of the MeSH term
``Observational Study'' in 2014.
We excluded publications that were also labeled as {\em Randomized Controlled Trial} or {\em
Clinical Trial}, leaving us with 176,336 exclusive observational studies.
We also removed papers that lacked either a structured abstract with a conclusion
subsection or an English full text, resulting in  125,388 observational studies
with a {\em Conclusion(s)} subsection and English full text. Finally, we
excluded articles whose conclusions did not contain any causal or correlational
claims—typically those focused solely on descriptive findings, recommendations,
study implications, or future work. This filtering process resulted in a final
dataset of 91,933 observational studies for analysis.
(see SI A.1 and SI Fig. S1 for details).

\noib{Identifying Causal and Correlational Claims}.
In our regression analysis, we defined the dependent variable as the presence or absence of a causal statement in
the conclusion of an observational study abstract---specifically, a conclusion is deemed causal
if it includes at least one sentence using causal language. 
To automatically classify sentences as causal, correlational, or neither, we
developed a causal language prediction model based on BioBERT
\citep{lee2019biobert}. Trained on a corpus of 3,061 annotated sentences, the
model achieved a macro-F1 score of 0.89 using 5-fold cross-validation
\citep{yu2019EMNLPCausalLanguage} (see SI B.1 for details).

Specifically, in our study, a {\em causal} statement uses terms indicating direct causation 
(e.g., {\em increase, make, lead to, effective in, contribute to}) 
or uses the modal verb {\em can} followed by a causal verb.
 A {\em correlational} statement is either 
a common {\em correlational} statement that employs language suggesting an association 
(e.g., {\em associated with, predictor, linked with}),
or a {\em conditional causal} statement that uses qualifiers (e.g., {\em
may, might, appear to, probably}) to tone down the level of certainty.
We classified the conditional causal statement into the correlational 
category, based on previous research
indicating that general readers often have difficulty distinguishing between these
two \citep{adams2017readers}.

\noib{Author Gender.}
We inferred gender from each author's forename, categorizing names as male,
female, or unknown. The ``unknown'' category was assigned when a forename was
missing or when the algorithm's confidence in the gender prediction was low.
Using a dataset of six million (name, gender) pairs from WikiData, we developed and open-sourced
a name-to-gender inference algorithm \cite{wang2025name2gender} based on LightGBM, a gradient boosting method \cite{ke2017LightGBMAH}.
The algorithm (see SI B.3)
uses individual letters from the beginning and end of each forename,
achieved performance comparable to the best of the five name-to-gender inference tools when
evaluated on a benchmark dataset \citep{Santamara2018ComparisonAB}.
Moreover, it offers the advantage of providing a confidence level for its gender predictions.

\noib{Author Country.}
Author affiliations in the PubMed metadata provide a basis for inferring country
information. However, this can be challenging when the country is not explicitly
mentioned (e.g., Beth Israel Deaconess Medical Center
is a well-known institution located in the United States, but the affiliation may not always specify USA or United States).
To address this, 
we developed an algorithm to infer an author’s country based on affiliation
metadata from each paper. This algorithm was trained on a dataset of
approximately 2.7 million organizations with known country information from the
ORCID public database (see SI B.2).
After obtaining country information for each available author affiliation, we aggregated these data
at the paper level. A paper was assigned to a specific country if all author
affiliations were classified under the same country and the average confidence
score exceeded 0.8; otherwise, it was labeled as “others.”

\noib{Author Name Disambiguation.}
An author’s writing experience is measured by the number of observational
studies they are associated with in our dataset.
However, it is challenging to determine which publications belong to which real-world authors,
especially when common names are shared by different authors or when the same author publishes under different names.
To uniquely identify authors, we used the Semantic Scholar API, which provides a
unique author ID for each researcher (details in SI A.3).
Semantic Scholar developed a machine learning-based system for author name disambiguation,
which has a performance comparable to the published state-of-the-art across eight benchmark datasets \cite{S2AND}.
Using their API, we were able to retrieve author IDs for 99.8\% of the authors in our dataset; for the remainder without an ID,
we assigned a unique random ID.

\noib{Random Effects.}
In our mixed-effects logistic regression analysis (see SI C), we incorporated three random
effects: 
(1) {\em Conclusion Length}, measured by the number of sentences, to account
for the greater likelihood of causal language in longer conclusions.
This adjustment is necessary given that the dependent variable is
defined as the presence of a causal statement in the conclusion of an
observational study abstract.
In our dataset, 99\% of conclusions contain four or fewer sentences.
(2) {\em Journal ISSN}, to capture variations in editorial guidelines and practices regarding
causal language across publications.
(3) {\em Publication Year}, to control for
temporal shifts in language use and evolving academic norms in response to
changing policies and cultural trends \cite{hyland2019}.

\noib{Logistic Regression Analysis.}
The results presented in Table~\ref{table__overall_effects} are based on a
logistic linear mixed-effects regression model (formula provided in
SI Table S3). Specifically, we used the R package $lme4$ \citep{Bates2014FittingLM} to do the regression analysis.

\section*{Code and Data}
The complete dataset of over 90,000 observational studies, along with the
associated analysis code and documentation, is available on
\href{https://github.com/junwang4/causal-language-use-in-observational-studies}{https://github.com/junwang4/causal-language-use-in-observational-studies}

\bibliographystyle{plainnat}
\bibliography{sect_9_references}

\begin{appendices}

\setcounter{section}{0}
\renewcommand{\thesection}{SI}
\renewcommand{\thesubsection}{\Alph{subsection}}

\renewcommand{\thetable}{S\arabic{table}}
\setcounter{table}{0}  
\renewcommand{\thefigure}{S\arabic{figure}}
\setcounter{figure}{0}

\section{Supplemental Material}

\subsection{Data Collection}

\subsubsection{Obtaining Observational Studies}

This study adopts the definition of {\em an observational study} provided by
the NIH National Library of Medicine\footnote{https://www.ncbi.nlm.nih.gov/mesh/68064888},
which states:
``[...] a work that reports on the results of a clinical study
in which participants may receive diagnostic, therapeutic, or other types of interventions,
but the investigator does not assign participants to specific interventions (as in an interventional study).''

We initially retrieved 180,520 papers from the PubMed database using the query
{\em Observational Study[Publication Type]} through the Entrez system
(see Fig.~\ref{fig__flowchart_data_collection} for the detailed procedure).
Most of these papers were published in 2014 or later, reflecting the adoption of the MeSH term Observational Study introduced in 2014.
We then excluded
publications that were also labeled as {\em Randomized Controlled Trial} or {\em
Clinical Trial}, leaving us with 176,336 exclusive observational studies. Next,
we removed papers that lacked either a structured abstract with a conclusion
subsection in the abstract or an English full text, resulting in 125,388 observational studies.
Finally, we excluded articles whose conclusions did not contain any causal or correlational
claims—typically those focused solely on descriptive findings, recommendations,
study implications, or future work. This filtering process resulted in a final
dataset of 91,933 observational studies for analysis.

\begin{figure}[!ht]
\centering
	\begin{tikzpicture}
    \tikzstyle{block} = [rectangle, rounded corners, draw, text width=28em, text centered, minimum height=3.5em]

    \node [block] (block1) {180,520 observational studies retrieved from the Entrez pubmed database
    using query {\tt Observational Study[Publication Type]} \\
    (up to 2025-08-01)};
    
    \node [block, below of=block1, yshift=-1.8cm] (block2) {
        176,336 exclusive observational studies
    };
    
    \node [block, below of=block2, yshift=-1.5cm] (block2_3) {
        134,686 have a structured abstract \\ 
    };
    \node [block, below of=block2_3, yshift=-1.5cm] (block3) {
        125,388 papers include a subsection of {\em Conclusion(s)} \\ in their {\em structured abstracts}
    };
    \node [block, below of=block3, yshift=-1.5cm] (block4) {
        118,711 are written in English in their full-text
    };
    \node [block, below of=block4, yshift=-1.5cm] (block5) {
        91,933 papers used in our regression analysis
    };

    \draw[->] (block1) -- (block2) node[anchor=west, midway, align=left] {
        Exclude papers that are labeled as \\ Randomized Controlled Trial or Clinical Trial
    };
    
    \draw[->] (block2) -- (block2_3) node[anchor=west, midway, align=left] {
        Exclude papers that do not have structured abstract 
    };

    \draw[->] (block2_3) -- (block3) node[anchor=west, midway, align=left] {
        Exclude papers that do not have a \\ conclusion subsection in the abstract
    };

    \draw[->] (block3) -- (block4) node[anchor=west, midway, align=left] {
        Exclude papers that are not written in English \\ in their full-text
    };

    \draw[->] (block4) -- (block5) node[anchor=west, midway, align=left] {
        Exclude papers that do not contain any causal or \\
        associational claims in the conclusion 
    };

\end{tikzpicture}
	\caption{Flowchart of the process for collecting observational studies for our analysis.}
	\label{fig__flowchart_data_collection}
\end{figure}

\subsubsection{Journal Rank}

We downloaded publicly accessible SciMago Journal Rank (SJR) data for 2024 and
prior years\footnote{https://www.scimagojr.com/journalrank.php}. For journals in
our dataset that could not be matched to the 2024 data via their ISSN, we
attempted to retrieve their information from previous years. If a journal
remained unmatched (which occurred for only 0.35\% of the papers), we assigned
it an SJR score of 0.49, corresponding to the bottom quartile in our list of
2,721 journals, based on the assumption that journals not listed in the SJR
database are likely to have a low rank.

\subsubsection{Author Writing Experience and Name Disambiguation}

An author’s writing experience is measured by the number of observational
studies they are associated with in our dataset.
However, it is challenging to determine which publications belong to which real-world authors,
especially when common names are shared by different authors or when the same author publishes under different names.
To disambiguate author names, we used the
Semantic Scholar API\footnote{https://api.semanticscholar.org/} to obtain a
unique ID for each author per article. 
Author IDs were successfully retrieved for 99.8\% of the authors; for the remainder, a unique random ID was generated and assigned.

Semantic Scholar's system of author name disambiguation, called S2AND (Semantic Scholar Author Name Disambiguation),
works by first grouping papers that share the same or lexically similar author name strings,
then using supervised learning to predict the likelihood that pairs of papers were written by the same individual 
based on features such as co-authors, affiliations, titles, venues, and abstracts, and finally applying unsupervised clustering 
to assign papers to distinct author identities based on the predicted pairwise similarities.
S2AND has demonstrated performance comparable to state-of-the-art methods across eight benchmark datasets \cite{S2AND}.

\subsection{Data Preprocessing}

\subsubsection{Causal and Correlational Claim Identification} \label{si_causal_or_correlational}

Building on the taxonomy introduced in \citep{sumner2014association,li2017nlp},
we annotated a corpus of 3,061 PubMed research conclusion sentences \citep{yu2019EMNLPCausalLanguage},
and developed a BioBERT-based transformer
model\footnote{https://github.com/junwang4/causal-language-use-in-science} to
classify a sentence in the conclusion subsection of an abstract into four levels
of claim strength: Causal, Conditional Causal, Correlational, or None. 
A causal claim uses terms indicating direct causation (e.g., increase,
make, lead to, effective in, contribute to) or uses the modal verb {\em can}
followed by a causal verb. A conditional claim uses qualifiers (e.g.,
may, might, appear to, probably) to tone down the level of certainty, whereas a
correlational claim employs language suggesting an association (e.g.,
associated with, predictor, linked with).

For this study, we merged the conditional causal and correlational
categories into a single correlational category, based on previous research
indicating that general readers often have difficulty distinguishing between these
two \citep{adams2017readers, bratton2020causal}. With this modification, we
fine-tuned the BioBERT-based model using the same training corpus and evaluation
procedure described in \citep{yu2019EMNLPCausalLanguage}, achieving a macro-F1
score of 0.89. Moreover, two empirical comparisons with the ChatGPT family of
models demonstrated that our specifically trained model still have advantages over ChatGPT for
this task \citep{kim2023can, chen2024evaluating}.

\begin{table}[ht]
    \centering
  \begin{tabular}{lrrrr}
  \toprule
    &               Precision &   Recall & F1-score  & Support \\
  \midrule
       causal &     0.878  &   0.844  &   0.861  &     494  \\  
    correlational  &    0.888  &   0.927  &   0.907  &    1211 \\
    none of them     & 0.919   &  0.895   &  0.907   &   1356 \\
      {\em Macro Average}     & 0.895 &    0.889 &     0.892 &      3061\\
  \bottomrule
  \end{tabular}
  \label{table__performance_of_causal_bert_model}
  \caption{Performance of a sentence-level causal and correlational claim detection model}
\end{table}

In our analysis, if a structured abstract’s conclusion subsection contains two
or more sentences (which occurs in 70.2\% of our dataset) and at least one
sentence is classified as causal, the entire conclusion is considered to include
a causal claim. Based on this criterion, 31.7\% of the 91,933 conclusions
contain a causal claim.

\subsubsection{Country Label} \label{si_country}

Author affiliations in the PubMed metadata provide a basis for inferring country
information. However, this can be challenging when the country is not explicitly
mentioned; for example, Beth Israel Deaconess Medical Center
is a well-known institution located in the United States,
but the affiliation may not always specify USA or United States.
requiring additional mapping or external knowledge to accurately assign country-level information.
To address this challenge, we developed a LinearSVC-based text classification model to infer the country
associated with an author’s affiliation, leveraging a training dataset of 2.7
million organizations with known country information from 
the ORCID database\footnote{https://support.orcid.org/hc/en-us/articles/360006897394-How-do-I-get-the-public-data-file}.

We evaluated the algorithm on 1,000 randomly sampled affiliations from the
PubMed metadata, which we group into two groups: (1) affiliations with
insufficient information to infer the country (e.g., affiliation listed simply
as “Division of Infectious Diseases”) and (2) affiliations with sufficient
information. For the small number of cases in Group 1 (7 affiliations), the
algorithm appropriately returned low confidence scores (below 0.8). In contrast,
for all affiliations in Group 2, the algorithm produced accurate country
inferences.  For additional details, please visit our GitHub
page\footnote{https://github.com/junwang4/affiliation-to-country-inference}.

Each paper is assigned to a country if all author affiliations belong to the
same country and the average confidence score exceeds 0.8. Otherwise, it is
categorized as “others.”
The “others” category includes 23,204 observational studies, broken down as
follows:
    (1) 18,399 from multiple countries,   
    (2) 2,336 from various other countries (each contributing fewer than 150 studies),
    (3) 1,636 with missing affiliation data, and
    (4) 833 with low-confidence classification (confidence score below 0.8).

\subsubsection{Gender Label} \label{si_gender}

Gender is inferred from an author's forename. Using a dataset of 6 million
(forename, gender) pairs from WikiData, 
we developed a Gradient Boosting-based algorithm that uses single-character
features extracted from fixed positions counted from the left or right side of
normalized forenames (up to 8 characters deep) to predict gender likelihood
that a name is male or female.

Our algorithm\footnote{https://github.com/junwang4/name-to-gender-inference}
performs comparably to the best of the five name-to-gender inference tools when
evaluated on a benchmark dataset of 5,779 manually labeled names
\citep{Santamara2018ComparisonAB}. In our implementation, a name is labeled as
male if its male confidence exceeds 0.82 and as female if its female confidence
exceeds 0.78; otherwise, it is classified as gender-unknown. These thresholds
were determined through searching a grid of possible values for best F1-score on the benchmark dataset under the following contraint:
at most 12.5\% of names are assigned to gender-unknown (this setting allows for about 98\% precision and 96\% recall for both genders).
Table~\ref{tab:gender_eval} summarizes the performance of our algorithm under these constraints.

\begin{table}[ht]
        \centering
        \begin{tabular}{c@{\hspace{.1in}}l}
            
        \begin{tabular}{lrrrr|cccc}
        \toprule
        \textbf{} & \textbf{F pred} & \textbf{M pred} & \textbf{U pred} & \textbf{Total} & \textbf{Precision} & \textbf{Recall} & \textbf{F1-score} & \textbf{Support} \\
        \midrule
        \textbf{F} & 1654 & 71 & (243) & 1968  & 0.976 & 0.959 & 0.968 & 1725 \\
        \textbf{M} & 40 & 3298 & (473) & 3811  & 0.979 & 0.988 & 0.983 & 3338 \\
        \toprule
        \end{tabular}
        \\ 
        \parbox[t]{5in}{
        (a) Results using constraint-optimized confidence thresholds: 0.82 for
        male and 0.78 for female predictions. The lower female threshold
        compensates for the training data's 75\% male majority, which biases
        predictions toward male.}
        \\ 
        \\

        \begin{tabular}{lrrrr|cccc}
        \toprule
        \textbf{} & \textbf{F pred} & \textbf{M pred} & \textbf{U pred} & \textbf{Total} & \textbf{Precision} & \textbf{Recall} & \textbf{F1-score} & \textbf{Support} \\
        \midrule
        \textbf{F} & 1641 & 79 & (248) & 1968 & 0.977 & 0.954 & 0.966 & 1720 \\
        \textbf{M} & 38 & 3320 & (453) & 3811 & 0.977 & 0.989 & 0.983 & 3358 \\
        \toprule
        \end{tabular}
        \\
        (b) Baseline results using symmetric confidence thresholds: 0.80 for both male and female predictions.
 \\ [10pt]
        \end{tabular}
        \caption{
Comparison of gender prediction performance under different confidence thresholds. 
Values in parentheses represent the number of uncertain predictions (labeled as "U pred").
}
        \label{tab:gender_eval}
\end{table}

\subsection{Logistic Linear Mixed-Effects Regression Analysis}

The results presented in Table 1 of the main text are based on a
logistic linear mixed-effects regression model (formula provided in
Table~\ref{tab_modelformula}). Specifically, we used the $glmer()$ function from
the R package $lme4$ \citep{Bates2014FittingLM} on 91,933 observations. In our
model, the dependent variable indicates whether the conclusion subsection of an
article's structured abstract contains at least one sentence with a causal
claim. The primary explanatory variables include: (1) Author writing experience
(measured as the number of observational studies published by the first/last
authors, log-transformed to address skewness); (2) Author gender (for first and
last authors); (3) Author country; and (4) Team size (the number of authors,
also log-transformed).


\newcommand{\ttt}{}

\begin{table}[ht]
    \centering
		\begin{tabular}{@{}ll@{ }ll@{}}
		\toprule
			\\ [-.1in]
        \multicolumn{4}{l}{\ttt Has\_a\_Causal\_Claim\_in\_Abstract\_Conclusion $\sim$} \\ [.2in]



        & \multicolumn{3}{l}{\ttt // VARIABLES OF INTEREST} \\ [.1in]
        &  & Num\_coauthors &   {\ttt // log-scaled}  \\
        & + & Num\_obs\_studies\_by\_{\bf first}\_author &  {\ttt// log-scaled} \\
        & + & Num\_obs\_studies\_by\_{\bf last}\_author &  {\ttt// log-scaled} \\
        & + & Gender\_of\_{\bf first}\_author & {\ttt // categorical: F, M, Unknown; reference: F} \\
        & + & Gender\_of\_{\bf last}\_author & {\ttt // categorical: F, M, Unknown; reference: F} \\
        & + & Country\_of\_authors & {\ttt// categorical: 40 exclusive countries and others; reference: others}\\ [.2in]

    & \multicolumn{3}{l}{\ttt // CONTROL VARIABLES} \\ [.1in]
        & + & Journal\_rank &   {\ttt// log-scaled: Scimago journal rank}  \\
        & + & Num\_obs\_studies\_in\_the\_journal &   {\ttt// log-scaled}  \\
        & +& $\sum_{i=1}^{N} \textrm{MeSH\_term}_i$  & {\ttt // binary: N$=$1400+ terms, each with 100+ papers; reference: absense} \\[.2in]

    & \multicolumn{3}{l}{\ttt // RANDOM EFFECT VARIABLES} \\ [.1in]
& + & (1 $|$ Year) & {\ttt // categorical: the majority are from 2013 to 2025} \\
& + & (1 $|$ Journal) & {\ttt// categorical: 2721 journals} \\  
& + & (1 $|$ Astract\_conclusion\_length) &  {\ttt// categorical: the majority are 1, 2, or 3 sentences}
\\ [.05in]
		\bottomrule
\end{tabular}
\caption{
The formula for the logistic linear mixed-effects regression model.
}
\label{tab_modelformula}
\end{table}

\subsection{Effects of Various Research Topics, Methods, and Concepts} \label{meshtermeffect}

Our dataset contains 1400+ Medical Subject Headings (MeSH) terms, each
treated as a control variable, with its absence serving as the reference
category. A MeSH term is included if it is assigned to at least 100 articles in
the dataset and is not a country name (since author country is analyzed
separately).  
Refer to the following GitHub repository for a detailed report on the effects of these MeSH terms. 
\href{https://github.com/junwang4/causal-language-use-in-observational-studies/blob/main/pdf/meshterm_effects.pdf}{https://github.com/junwang4/causal-language-use-in-observational-studies}

\end{appendices}

\end{document}